\documentclass{article}
\usepackage[preprint]{spconf}
\usepackage{amsmath,graphicx}
\usepackage{fancyhdr}

\makeatletter
\def\thickhline{%
  \noalign{\ifnum0=`}\fi\hrule \@height \thickarrayrulewidth \futurelet
   \reserved@a\@xthickhline}
\def\@xthickhline{\ifx\reserved@a\thickhline
               \vskip\doublerulesep
               \vskip-\thickarrayrulewidth
             \fi
      \ifnum0=`{\fi}}
\makeatother

\newlength{\thickarrayrulewidth}
\title{Welsch Based Multiview Disparity Estimation}
\name{James L. Gray, Aous T. Naman, David S. Taubman}
\address{School of Electrical Engineering and Telecommunications \\
        Faculty of Engineering\\
        University of New South Wales}
\usepackage{tikz}
\usepackage{textcomp}
\usepackage{hyperref}
\usepackage{lipsum}

\begin{document}
\ninept
\maketitle
\thispagestyle{fancy}
\pagenumbering{gobble}
\begin{abstract}
In this work, we explore disparity estimation from a high number of views. We experimentally identify occlusions as a key challenge for disparity estimation for applications with high numbers of views. In particular, occlusions can actually result in a degradation in accuracy as more views are added to a dataset. We propose the use of a Welsch loss function for the data term in a global variational framework for disparity estimation. We also propose a disciplined warping strategy and a progressive inclusion of views strategy that can reduce the need for coarse to fine strategies that discard high spatial frequency components from the early iterations. Experimental results demonstrate that the proposed approach produces superior and/or more robust estimates than other conventional variational approaches.
\end{abstract}
\section{Introduction}
\label{sec:intro}
Disparity estimation has been studied extensively in the literature and many different approaches have been taken for this problem. Most existing approaches focus on stereo pairs and excellent results have been achieved in this context \cite{Scharstein2001} \cite{yang2019hierarchical} \cite{song2020edgestereo} \cite{Xu2020}. However, the intuitive assumption that adding more views of a given scene will make disparity estimation more accurate does not actually hold. In scenes with occlusions, it turns out that as more views are added, with larger baselines, the increased areas of occlusion pose problems for most disparity estimation algorithms \cite{Kang2001}. This is unfortunate, since larger baselines allow for greater precision in the estimated disparity \cite{okutomi1993multiple}. 

Current hand-crafted methods of multiview disparity estimation with large numbers of views generally attempt to deal with this by using explicit occlusion based reasoning. This is done to reduce the influence of occluded areas on the estimated disparity. These techniques often use heuristics such as confidence based measures \cite{Strecha2002} or specific visibility reasoning  \cite{Kang2001} \cite{Zhu2015} \cite{Hu2012}. These are often based on preliminary estimates of the disparity field themselves, so that early errors can potentially be reinforced. Other approaches treat occluded areas simply as outliers \cite{Bailer2012} \cite{10.1007/978-3-540-88682-2_58} \cite{Jeon2015} and reduce their effect on the solution. This is conceptually similar to the use of robust cost functions in optical flow, reducing the influence of outliers on the solution without explicitly identifying them. Common robust cost functions include the Huber norm \cite{shulman1989regularization} and similar $L^1$-like norms \cite{brox2004high} \cite{volz2011modeling} the Geman and McClure norm, the Lorentzian norm \cite{Black1996}, the Tukey norm \cite{odobez1995robust} and the Welsch loss function \cite{Holland1977}.

In this paper we experimentally show that occlusions are a key problem for extending stereo disparity to a large number of views. We also propose a novel method of dealing with these occlusions in a multiview context using a Welsch loss function based data term. In particular, we propose an automatic selection process for Welsch loss parameter $\sigma_d$, a progressive approach to including the multiple views and a warping strategy that uses a disciplined multiple hypotheses method for upsampling the disparity field. We evaluate the proposed method using a synthetic dataset with 31 views of the same scene as well as a 4D Light Field Benchmark training dataset \cite{Honauer2016}.

This paper is organised as follows. The proposed Welsch-$L^1$ multiview disparity algorithm is detailed in Section \ref{sec:algorithm}. The progressive inclusion of views is detailed in Section \ref{sec:ProgViews}. The warping strategy which uses a disciplined method for upsampling disparity fields is discussed in Section \ref{sssec:WarpStrat}.
The performance of the proposed algorithm is evaluated in Section \ref{sec:Eval}. Conclusions are provided in Section \ref{sec:Conclusion}.

\section{Welsch-$L^1$ Disparity Estimation} \label{sec:algorithm}

For simplicity, we assume that we are dealing with images taken from cameras that are co-planar, with identical orientation. This means that the two dimensional disparity vector $\mathbf{d}_{p,q}(\mathbf{s})$, describing the apparent displacement between location $\mathbf{s}$ in view $p$ and a corresponding location in view $q$, can be written as \vspace{-1mm}
\begin{equation} \label{eqn:disp->reciprocal_depth} 
    \mathbf{d}_{p,q}(\mathbf{s}) = \mathbf{B}_{p,q}\cdot F\cdot r(\mathbf{s})
    \vspace{-1.5mm}
\end{equation}
where $F$ is the focal length, $r(\mathbf{s})$ is the reciprocal depth or normalised disparity, and $\mathbf{B}_{p,q}$ is the baseline between views $p$ and $q$. It is more convenient to write this as \vspace{-1mm}
\begin{equation} 
    \mathbf{d}_{p,q}(\mathbf{s}) = \mathbf{B}^\prime_{p,q}\cdot w(\mathbf{s})
    \vspace{-1.5mm}
\end{equation}
where $\mathbf{B}^\prime$ and $w(\mathbf{s})$ are normalized baselines and reciprocal depth fields such that the nearest pair of views have $\mathbf{B}^\prime \in \{0,1\}^2$.

We apply a global variational approach, similar to \cite{horn1981determining} to minimise the energy of \vspace{-1.5mm}
\begin{equation}\vspace{-1mm}
    E = E_{data} + \alpha^2 E_{reg},
\end{equation}
where $E_{data}$ is the data term, $E_{reg}$ is the regularisation term and $\alpha$ determines the relative significance of the regularisation term.
\vspace{-2mm}
\subsection{Data term}
The use of the Welsch loss function has been shown previously to be highly effective in limiting the influence of outliers and occlusions on a flow field in the domain of optical flow \cite{Young2019}. In the domain of multiview disparity estimation with potentially large occluded areas, this turns out to be particularly valuable. We therefore define our data term as \vspace{-1.5mm}
\begin{equation} \label{eqn:DataTerm} \vspace{-1.5mm}
    E_{data} = \int_{\Omega} \sum_p \phi_{\sigma_d}(I_q(\mathbf{s}) - I_p(\mathbf{s} + \mathbf{B}^\prime_{p, q} w(\mathbf{s}))) d \mathbf{s},
\end{equation}
where $\Omega$ is the image domain, $q$ is set to be the single reference view, and $\phi_{\sigma_d}$ is the Welsch loss function with the parameter $\sigma_d$. This is defined as
\begin{equation}
    \phi_{\sigma_d}(x) = \sigma_d ^2 (1 - \exp(-x^2 / 2 \sigma_d ^2)).
\end{equation}
%
% The selection of $\sigma_d$ will be discussed later in \ref{sssec:select_sig_d}. 

For small $w(\mathbf{s})$, equation \eqref{eqn:DataTerm} can be linearised as \vspace{-0.5mm}
\begin{equation} \label{eqn:LinDataTerm}
    E_{data} = \int_\Omega \sum_p \phi_{\sigma_d}(\langle\nabla I(\mathbf{s}), \mathbf{B'}_{p,q}\rangle w(\mathbf{s}) + \delta I_{p, q}(\mathbf{s})) d \mathbf{s}, \vspace{-1mm}
\end{equation}
where $\langle \nabla I(\mathbf{s}) , \mathbf{B}^\prime_{p,q} \rangle$ is the inner product of the gradient $\nabla I$ and the normalised baseline vector $\mathbf{B}^\prime_{p,q}$ and $\delta I_{p, q}(\mathbf{s})$ is the difference in image intensity between $I_q(\mathbf{s})$ and $I_p(\mathbf{s})$. For larger $w(\mathbf{s})$ values, a warping strategy is required and this is discussed in Section \ref{sssec:WarpStrat}.
\vspace{-2mm}
\subsection{Regularisation term}
This paper primarily focuses on the data term, because the data term is what primarily changes when extra views are added. As shown in equation \eqref{eqn:RegTerm} below, we adopt the $L^1$ norm for our regularisation term, as a reasonable choice which does not introduce additional parameters that might complicate the analysis. Specifically
\begin{equation}\label{eqn:RegTerm}
    E_{reg} = \int_\Omega \psi(\|\nabla w(\mathbf{s})\|)d\mathbf{s}, \vspace{-2mm}
\end{equation}
where $\psi(x) = x$.
\vspace{-2mm}

% We have selected the Huber function because it is convex and thus all local minima are global minima \cite{Shulman1989}. Additionally, it approximates the $L^1$ norm when $\delta$ is small. This is significantly more robust than a least squares approach, since up to half of the observations can be outliers before the estimate is impacted \cite{lopuhaa1991breakdown}. It is also highly similar to regularisation approaches that use the modified $L^1$ norm $\rho(x) = \sqrt{x^2 + \delta^2}$ \cite{brox2004high} \cite{volz2011modeling}

% where $\rho_H$ is the Huber function:
% \begin{equation}
%     \psi_H(x) = \begin{cases}
%      \frac{1}{2\delta}x^2, &\text{ for } |x| \leq  \delta \\
%     |x| - \frac{1}{2}\delta, &\text{ otherwise.}
%     \end{cases}
% \end{equation}
% We select $\delta = 0.0001$. We have selected the Huber function because it is convex and thus all local minima are global minima \cite{Shulman1989}. Additionally, it approximates the $L^1$ norm when $\delta$ is small. This is significantly more robust than a least squares approach, since up to half of the observations can be outliers before the estimate is impacted \cite{lopuhaa1991breakdown}. It is also highly similar to regularisation approaches that use the modified $L^1$ norm $\rho(x) = \sqrt{x^2 + \delta^2}$ \cite{brox2004high} \cite{volz2011modeling}.

\subsection{Optimisation}

To discuss our optimisation approach, we must introduce the numerical approximations we use in the energy function. To approximate a continuous $\nabla I$ we use Derivative of Gaussian filters in both the $x$ and $y$ directions ($G_{x, \sigma}, G_{y, \sigma}$) with $\sigma = 0.75$. We write this as \vspace{-1mm}
\begin{equation}
    \nabla I(\mathbf{s}) \approx  \left[(G_{x, \sigma}  * (I_q + I_p))(\mathbf{s}) , (G_{y, \sigma} * (I_q + I_p))(\mathbf{s})\right].
    \vspace{-1mm}
\end{equation}
This is equivalent to sampling the derivative of Gaussian blurred image, $G_\sigma(\mathbf{s}) * (I_p(\mathbf{s}) + I_q(\mathbf{s}))$. Thus, we avoid aliasing associated with using discrete derivative operators.

Similarly, $\delta I_{p, q}$ is calculated by subtracting one image from the other and then  blurring the image pair with a Gaussian filter $G_\sigma$, \vspace{-1mm}
\begin{equation}
    \delta I_{p, q}(\mathbf{s}) \approx G_\sigma(\mathbf{s}) * (I_{q}(\mathbf{s}) - I_{p}(\mathbf{s})).
    \vspace{-1mm}
\end{equation}

We use an Iteratively Reweighted Least Squares approach to minimise the Energy function. At each stage we use a conjugate gradient technique to solve the weighted Euler-Lagrange equation
\begin{equation}
\begin{split}
    \label{eqn:EL-eqn}
    W_d(\mathbf{s}) \langle\nabla I(\mathbf{s}), \mathbf{B'}_{p,q}\rangle ^2  w(\mathbf{s}) + W_d(\mathbf{s})\delta I_{p,q} \langle\nabla I(\mathbf{s}), \mathbf{B'}_{p,q}\rangle \\= W_r(\mathbf{s}) \alpha ^ 2 \nabla ^2 w(\mathbf{s}),
\end{split}
\end{equation}
where $W_d(\mathbf{s})$ and $W_r(\mathbf{s})$ are the data term and regularisation term weights and $\nabla^2 w(\mathbf{s})$ is approximated using the discrete point spread function 
\begin{equation} \label{eqn:nabla->w}
    \begin{bmatrix}
        1/12 & 1/6 & 1/12 \\
        1/6 & -1 & 1/6 \\
        1/12 & 1/6 & 1/12
    \end{bmatrix}. 
\end{equation}

\section{Progressive Inclusion of Views} \label{sec:ProgViews}

Disparity is proportional to the baseline as per equation \eqref{eqn:disp->reciprocal_depth}. If the outer views are considered initially, we may find that the disparity field $\mathbf{d}(\mathbf{s})$ is too large. This results in the linearisation in equation \eqref{eqn:LinDataTerm} becoming inaccurate as we will explain shortly. To prevent this, we propose an approach analogous to a coarse-to-fine strategy. We start with a subset of nearby views and then progressively include further away views, until all views are used. This allows the technique to include the important high resolution disparity information from the further away views, whilst preventing issues associated with large baselines from becoming a significant.

\begin{figure}
    \centering
    \includegraphics[width=7.5cm]{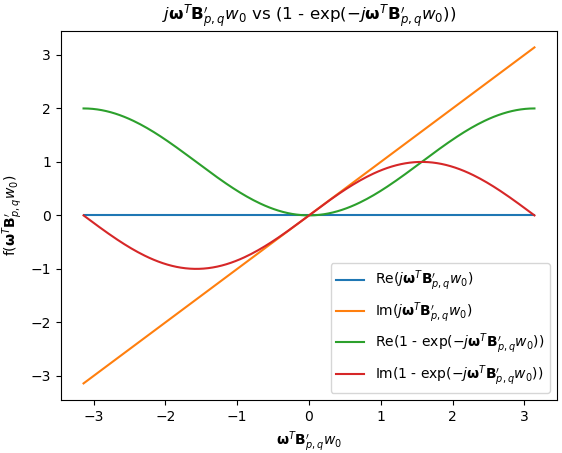}
    \vspace{-2.5mm}
    \caption{$j\boldsymbol{\omega} ^ T \mathbf{B}^\prime_{p, q} w_0$ plotted alongside $(1 - e^{-j\boldsymbol{\omega} ^ T \mathbf{B}^\prime_{p, q} w_0})$.}
    \label{fig:approx_big}
    \vspace{-5mm}
\end{figure}

The linearisation in equation \eqref{eqn:LinDataTerm} is based on a Taylor series approximation of a disparity shift. To simplify matters we consider a constant normalised reciprocal depth ${w}_0$ with the shift performed on the reference view $I_q(\mathbf{s})$ so that, 
\vspace{-1mm}
\begin{equation} \label{eqn:space_domain_approx} 
    I_q(\mathbf{s}) - I_q(\mathbf{s} + \mathbf{B}^\prime_{p, q} w_0)) \approx \langle \nabla I_q(\mathbf{s}), \mathbf{B}^\prime_{p, q}  \rangle w_0. \vspace{-1mm}
\end{equation}
We can take the Fourier transform of \eqref{eqn:space_domain_approx} as
\begin{equation} \label{eqn:Fourier_domain_approx} \vspace{-1mm}
    \hat{I}_q(\boldsymbol{\omega}) - \hat{I}_q(\boldsymbol{\omega})e^{j\boldsymbol{\omega}^T\mathbf{B}^\prime_{p, q} w_0} \approx \hat{I}_q(\boldsymbol{\omega})j\boldsymbol{\omega}^T\mathbf{B}^\prime_{p, q} w_0, \vspace{-1mm}
\end{equation}
where $\hat{I}_q$ is the Fourier transform of $I_q$ and 
he approximation in equation \eqref{eqn:Fourier_domain_approx} relies on \vspace{-1mm}
\begin{equation} 
    1 - e^{j\boldsymbol{\omega}^T\mathbf{B}^\prime_{p, q} w_0} \approx j\boldsymbol{\omega}^T\mathbf{B}^\prime_{p, q} w_0.
    \vspace{-1mm}
\end{equation}
As illustrated in Figure \ref{fig:approx_big}, when $\boldsymbol{\omega}^T\mathbf{B}^\prime_{p, q} w_0$ increases beyond $\pi/2$ the approximation degrades. 

One way of solving this problem is to use a coarse-to-fine strategy with spatial low-pass filtering \cite{brox2004high} \cite{Meinhardt-Llopis2013}. This ensures that $\hat{I}_q(\boldsymbol{\omega})$ is sufficiently bandlimited, effectively limiting $\boldsymbol{\omega}$. If done correctly, $\boldsymbol{\omega}^T\mathbf{B}^\prime_{p, q} w_0$ is kept small. Where required, such as where the disparities between adjacent views are too large, we apply this strategy. 

Although spatial low-pass filtering is sometimes unavoidable, it has the strong disadvantage of discarding high frequency details that may be required to reduce the ambiguity of correspondences.  For this reason, our proposed approach relies primarily on limiting $\mathbf{B}^\prime_{p, q}$, rather than $\mathbf{\omega}$, by using views close to the reference view and then progressively increasing
the number of views.  At stage $s$, where $s \in {0, 1, 2, ... S}$, we include a view
$p$ if $\Vert \mathbf{B}^\prime_{p,q} \Vert_\infty \leq k + s c $, where $k$ and $c$ are constants. At each stage $s$, we perform IRLS optimisation to obtain an estimated disparity. We warp all views according to the estimated disparity, to ensure that at $s+1$, the error in the approximation in equation \eqref{eqn:Fourier_domain_approx} is rather small. 

It would be possible to extend this algorithm to use multiple different reference views when estimating the disparity. However, this would require fusing several depth estimates associated with different reference views and is therefore beyond the scope of this paper.

\section{Warping strategy} \label{sssec:WarpStrat}

To minimise the introduction of aliasing during warping, which is particularly problematic around strong discontinuities, we perform warping at twice the resolution of the images. We upsample the image  using windowed-sinc interpolation. To similarly increase the resolution of the disparity we use a disciplined disparity upsampling approach.

The intuitive approach to upsampling a disparity field is to treat it the same as an image. Unlike image data, however, there is no reason to believe that disparity fields should exhibit bandlimited sampling properties. Instead of bandlimited interpolation, therefore, we propose a multi-hypothesis approach.

Specifically, we first use a nearest-neighbour policy to upsample the discrete disparity field by a factor of $2$ and then hypothesize $9$ high resolution disparity fields by applying unit shifts to the upsampled field. We can write this as
\vspace{-1.5mm}
\begin{equation} 
    w_{\mathbf{h}}[\mathbf{m}] =
    w_{\uparrow 2}[\mathbf{m}+\mathbf{h}],
    \vspace{-1.5mm}
\end{equation}
where $\mathbf{h} \in \{0,\pm 1\}^2$, $w_{\uparrow 2}$ is the initial upsampled disparity, and $w_{\mathbf{h}}$ is the candidate corresponding to shift $\mathbf{h}$.

We then warp the upsampled image, which is inherently smooth, using each of these disparity fields to produce 9 candidate images. These images are then averaged.  This can be represented as \vspace{-1.5mm}
\begin{equation}\vspace{-1mm}
    I_{\uparrow, \mathcal{W}_d}[x, y] = \frac{1}{9} \sum^1_{h_y=-1} \sum^1_{h_x=-1} \mathcal{W}_{d_\mathbf{h}}(I_\uparrow)[x, y],\vspace{-1mm}
\end{equation}
where $d_\mathbf{h} = \mathbf{B}^\prime w_{\mathbf{h}}$, $I_\uparrow$ is the upsampled image and $\mathcal{W}_{d_\mathbf{h}}()$ is the bilinear warping operator. $I_{\uparrow, \mathcal{W}_d}$, is then decimated using windowed-sinc filters to produce our final warped image.

% \subsection{Multi resolution framework}

% Where required, we use a multiscale strategy when disparities are expected to be larger than one pixel between adjacent views. We use similar strategy to Meinhardt-Llopis et al. where we downsample repeatedly by a factor of 1.5 until the number desired resolution is reached \cite{Meinhardt-Llopis2013}. To do this, we use a Gaussian filter of $\sigma=0.5$ to remove the high frequency components of the image. Then we use first order spline interpolation for the downsampling of fractional pixels. 

\section{Evaluation}\label{sec:Eval}

% \subsubsection{Selection of $\sigma_d$} \label{sssec:select_sig_d}

To fully implement our proposed algorithm, we select $\sigma_d$ by first estimating the squared error in the optical flow equation across the image domain for each view adjacent to the centre view. 
% We use a mesh-based warping method proposed by Li et al. with 1 pixel by 1 pixel triangles, to identify occluded areas and remove those from the set of pixels which are averaged over \cite{Li2019}. 
Specifically, 
\begin{equation}
    \sigma_d = \frac{1}{|\mathcal{I}_c|}\sum_p \sqrt{\frac{1}{|\Omega_p|}\sum_{\Omega_p} (\langle \nabla I(\mathbf{s}), \mathbf{B}_{p,q}^\prime \rangle w^{(k)}{(\mathbf{s})} + \delta I_{p, q}) ^ 2},
\end{equation}
where, $\mathcal{I}$ is the set of all views and $\mathcal{I}_c$ is the set of views adjacent to the centre view. We only use centre views because they are likely to have the smallest occluded areas. We calculate $\sigma_d$ each time we perform a reweighting step during IRLS optimisation. 

We also impose a monotonicity constraint, explicitly preventing any increase in $\sigma_d$ as the iteration proceeds. This is reasonable, because we expect the solution for $w$ to become increasingly accurate with reweighting.

In order to realise $L^1$ cost functions in the IRLS optimisation framework, we actually employ a Huber function with linear transition point $\epsilon=0.0001$.  This is used for the $L^1$ regularisation term, and when comparing our propose Welsh loss data term with $L^1$.
% 
% In order to realise  $L^1$ implementation in an IRLS optimisation framework, we use the Huber function as the cost function for the regularisation term. We select a $\epsilon = 0.0001$ for the Huber function.

For comparison purposes, to ensure that the regularisation strength $\alpha$ has roughly the same weight across increasing numbers of views, we scale $\alpha$ by $\sqrt{|\mathcal{I}|}$.

% We then compare each one to the ground-truth disparity field to obtain a result. One can use the root mean square error or any other kind of error metric with this method of upsampling. In this paper, we use the root mean square error can be found using the equation:
% \begin{equation}
%     RMSE = \sqrt{\frac{1}{9N}\sum_k \sum_n (r_{est_k}[n] - r_{gt}[n])^2}
% \end{equation}

% Another method we propose is to evaluate the prediction error, but exclude occluded regions... This is achieved using a mesh-based warping method which identifies occluded regions. This can be be done forward or in the reverse direction. Using the equation:
% \begin{equation}
%     P_{RMSE} = \sqrt{\frac{1}{|N^*|}\sum_{n \in N^*} (I_{tgt}[n] - W * I_{ref}[n])^2}
% \end{equation}
% where $N^*$ is the set of all pixels that are not occluded, $I_{ref}$ is the view which is warped and $I_{tgt}$ is the view which it is compared to.

% Occluded areas are removed from this comparison because these areas cannot be predicted from the reference view.

\begin{figure}[htb]
\begin{minipage}[b]{.48\linewidth}
    \centering
    \centerline{\includegraphics[width=4cm]{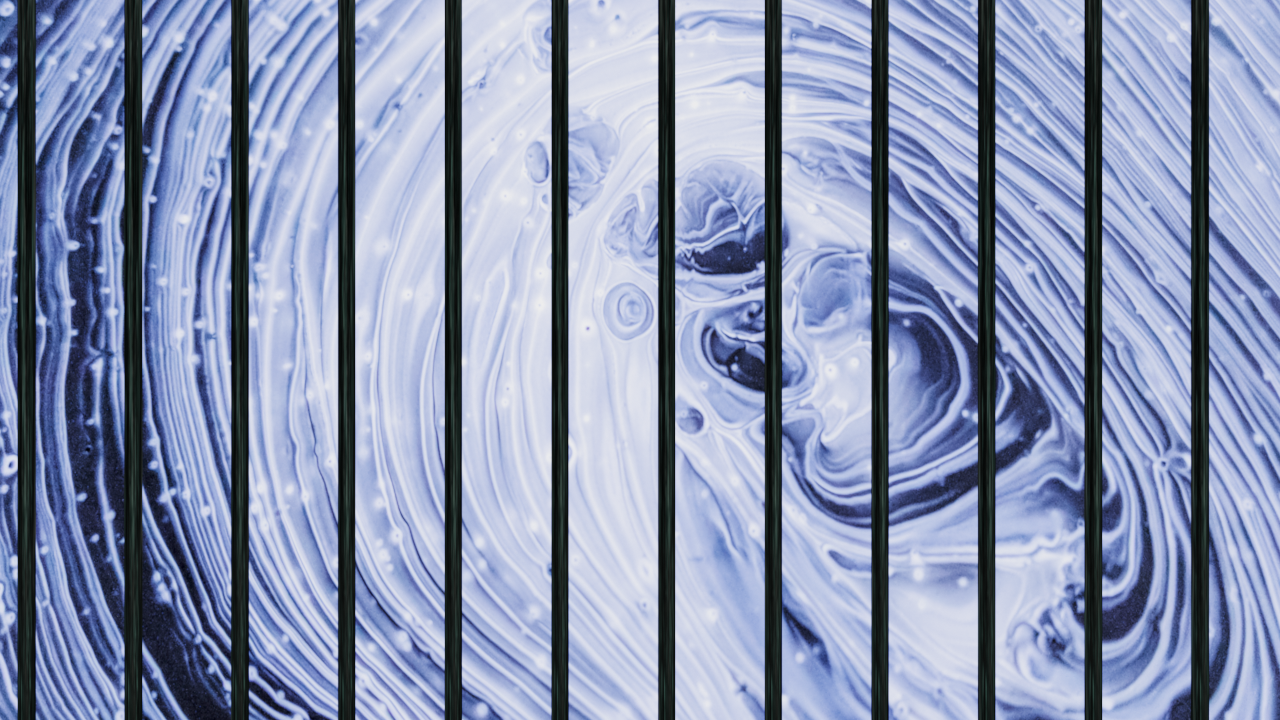}}
    \centerline{(a)}\medskip
    
\end{minipage}
\hfill
\begin{minipage}[b]{.48\linewidth}
    \centering
    \centerline{\includegraphics[width=4cm]{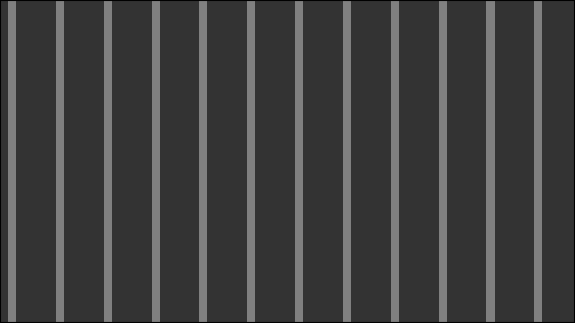}}
    \centerline{(b)} \medskip
\end{minipage}
\vspace{-5mm}
    \caption{The centre view in our dataset (a) and the corresponding Ground Truth disparity field (b).}
    \label{fig:Cam00}
\vspace{-2.5mm}
\end{figure}

We first use synthetic data to evaluate our algorithm using a dataset involving 31 views of the same scene. The spacing between views is 1.25mm and they are 360 pixels tall and 640 pixels wide. The views are arranged along the $x$-axis. The focal length of the cameras is 50mm. We select the small spacing to avoid requiring a multi-resolution approach on this dataset. To simulate a real imaging process, we add Gaussian white noise of variance $\sigma_n ^2 = 0.01$ to the images, where image intensity values are real and have the range $[0, 1]$.  
% Non-lambertian reflection is not considered in the simulation process. 
The centre view of this dataset is  shown in Figure \ref{fig:Cam00} (a) and the Ground Truth disparity field is shown in Figure \ref{fig:Cam00} (b). Note that the vertical bars produce obvious occlusions.

% We downsample the images by a factor of 2 prior to applying our approach, to ensure that any high frequency artefacts due to aliasing present do not cause problems for our experiment.
The ground truth disparity field is twice the resolution of the estimated disparity field, so we use our disciplined method of upsampling disparity to evaluate our results, (see Section \ref{sssec:WarpStrat}). We use the root mean square error, but we  include the 9 different upsampled disparity hypotheses (indexed by $m$)
\begin{equation}
    RMSE = \sqrt{\frac{1}{9|\Omega|}\sum_m \sum_n (r_{est_m}[n] - r_{gt}[n])^2}.
\end{equation}

\begin{figure}[htb]
    \vspace{-1mm}
    \centering
    \centerline{\includegraphics[width=7.5cm]{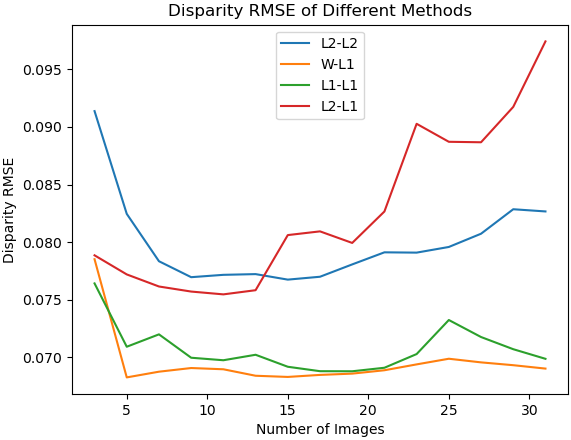}}
    \vspace{-2.5mm}
    \caption{RMSE results for the different approaches. $L^2$-based results have increasing RMSE values from 10 views onwards. For the Welsch ($W$-$L^1$) and the $L^1$-$L^1$ approach, this does not occur. For large numbers of views the Welsch based method worked best.}
    \label{fig:ResGraph}
    \vspace{-2.5mm}
\end{figure}

We denote our proposed model with a Welsch cost function for the data term and $L^1$ regularisation as Welsch-$L^1$. We compare it with an $L^2$-$L^2$ approach, an $L^2$-$L^1$ approach and an $L^1$-$L^1$ approach. The first norm refers to the cost function used for the data term, $\phi()$, and the second refers to the cost function used for the regularisation term, $\psi()$. For all three models, we use our progressive inclusion of views approach beginning with the 3 middle views and adding 2 more views at each stage. We chose the $\alpha$ values $0.01, 0.025, 0.05, 0.1, 0.2, 0.5, 1, 2, 5$ and ran each algorithm with each $\alpha$ value. The plots for a given approach in Figure \ref{fig:ResGraph} show the best error value for a given number of images over all $\alpha$ for that approach.  This is necessary, because the optimum $\alpha$ value is generally different for each algorithm. 

% Need to re do this section when the results are done.
Figure \ref{fig:ResGraph} shows that all $L^2$-based approaches have error levels that trend downward initially and then upward after roughly 10 views as more views are added. This is because in the initial stage, the noise in the images is uncorrelated so adding more images makes the data term more accurate. However, as we add views further and further apart we increase the size of occluded areas and with an $L^2$ data term this increasingly pollutes the disparity field. This is because occluded and non-occluded areas contribute with equal weight, pulling the solution towards their average. 

\begin{figure}
\begin{minipage}[b]{.48\linewidth}
    \centering
    \centerline{\includegraphics[width=3cm]{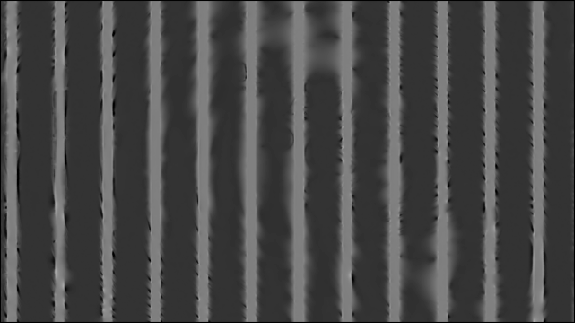}}
    \centerline{(a) $L^2$-$L^2$ disparity field}\medskip
\end{minipage}
\hfill
\begin{minipage}[b]{0.48\linewidth}
    \centering
    \centerline{\includegraphics[width=3cm]{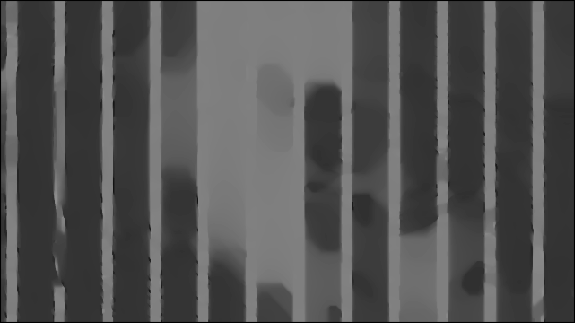}}
    \centerline{(b) $L^2$-$L^1$ disparity field}\medskip
\end{minipage}

\begin{minipage}[b]{0.48\linewidth}
    \centering
    \centerline{\includegraphics[width=3cm]{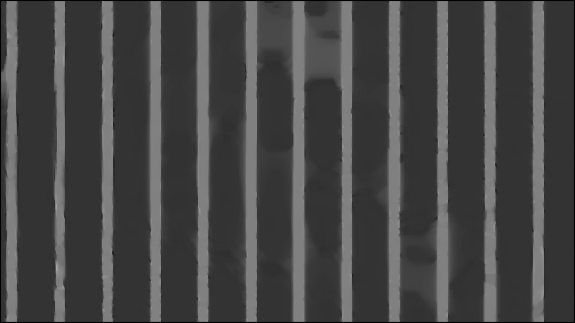}}
    \centerline{(c) $L^1$-$L^1$ disparity field}\medskip
\end{minipage}
\hfill
\begin{minipage}[b]{0.48\linewidth}
    \centering
    \centerline{\includegraphics[width=3cm]{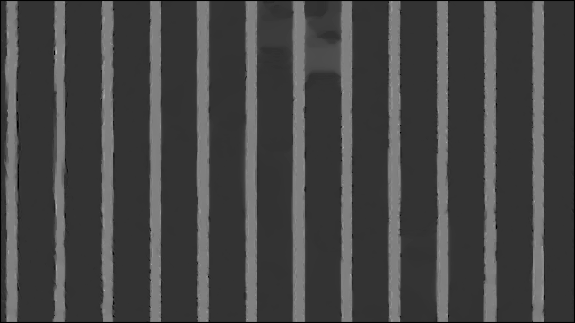}}
    \centerline{(d) Welsch-$L^1$ disparity field}\medskip
    \label{fig:ResPics}
\end{minipage}
\vspace{-4.5mm}
\caption{The best estimated disparity fields obtained with each algorithm on all 31 views.}
% \vspace{-4mm}
\label{fig:EstDisparitySlats}
\end{figure}

With the $L^1$-$L^1$ norm, we see the initial downward trend, but no increase after roughly 10 views. We see similar, but improved results for the Welsch-$L^1$ approach compared to the $L^1$-$L^1$ results. We can see from Figure \ref{fig:EstDisparitySlats} (c) and (d) that the Welsch data term is better at determining where the background begins and ends when compared with the $L^1$ data term.

% The different regularisation strategies also had key effects on these results. The $L^1$ regularised results had sharper discontinuities than the $L^2$ regularised results, as seen in Figure \ref{fig:EstDisparitySlats}. This is expected to be the case because an $L^1$ regulariser favours piece-wise-constant flow fields rather than smooth flow fields. % It is interesting to note that the $L^1$ method also is prone to staircasing artefacts. These haven't been observed in this dataset, but they have been well documented in the literature \cite{FORTUN20151}. 

\begin{figure}[htb]
    \centering
    \vspace{-1mm}
    \centerline{\includegraphics[width=7.5cm]{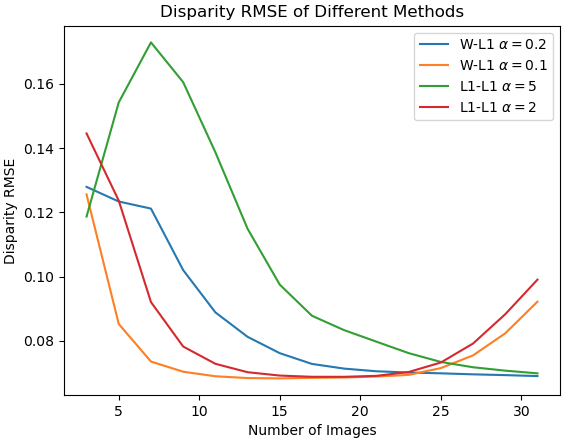}}
    \vspace{-2.5mm}
    \caption{RMSE results for the Welsch-$L^1$ and $L^1$-$L^1$ for a few values for $\alpha$. Note that the $L^1$-$L^1$ algorithm provides good results for a smaller number of views than the Welsch-$L^1$ algorithm. At present, we do not have an explanation for the worsening of the $L^1$-$L^1$ results for $\alpha = 5$ up to around 7 to 9 views.}
    \label{fig:Welsch_vs_L1_slats}
    \vspace{-5mm}
\end{figure}

We also note that the $L^1$-$L^1$ lower bound involves results from four $\alpha$ values, while the Welsch-$L^1$ lower bound involves results from three. In Figure \ref{fig:Welsch_vs_L1_slats} we see that the $L^1$-$L^1$ results for a given $\alpha$ are less consistent than the Welsch-$L^1$ results. This is particularly important, since in real applications one has no way to know the best performing regularisation strength $\alpha$.

We also evaluate the performance of the proposed approach using a 4D Light Field Benchmark training dataset \cite{Honauer2016}. We compare the estimated disparity from our algorithm with the ground truth data which has a resolution that is the same as that of the image. We use $\alpha =$ 0.01, 0.025, 0.05, 0.1, 0.2, 0.5, 1, 2, 5 for all four approaches. 

\begin{figure}
    \centering
    \centerline{\includegraphics[width=7.5cm]{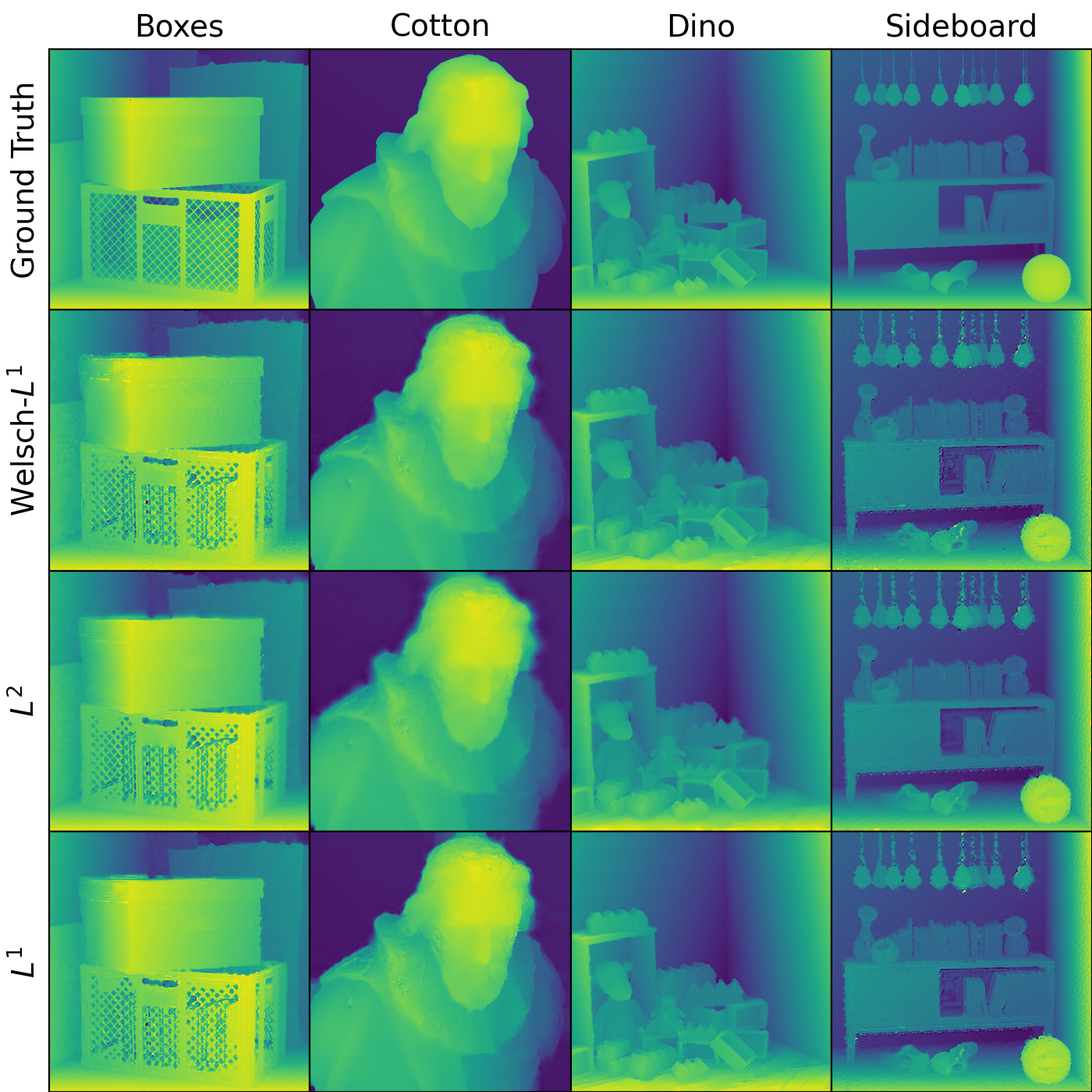}}
    \vspace{-2.5mm}
    \caption{Ground truth and estimated disparity fields with 17 views of four scenes from the 4D Light Field Training Dataset \cite{Honauer2016}. The colour scale is shared between the plots of the same scene, but not between scenes.}
    % \vspace{-2.5mm}
    \label{fig:4DLightField}
\end{figure}

Because the ground truth field is the same resolution as the images, we simply use RMSE to evaluate the performance of our algorithm. Our progressive inclusion of views strategy begins with selecting the central view and a neighbouring view. We then add views one a time, vertically and horizontally only, leading to a crosshair camera array. At each stage we produce a disparity estimate. In total, we use 17 of the 81 views. In Table \ref{tab:HonauerTrainingResults} we detail the RMSE results for our four algorithms on each of the training scenes and we show the disparity fields in Figure \ref{fig:4DLightField}.

\begin{table}[htb]
% \vspace{-2.5mm}
    \centering
    \begin{tabular}{|c | c c c c|}
        \thickhline
        \textbf{Scene} & $L^2$-$L^2$ & $L^2$-$L^1$ & $L^1$-$L^1$  & Welsch-$L^1$ \\
        \thickhline
        % Boxes Overall & 0.335 & 0.409 & {0.321} & \textbf{0.319} \\
        % \hline
        Boxes & 0.349 & 0.446 & \textbf{0.309} & 0.317 \\
        \hline
        % Cotton Overall & 0.303 & 0.358 & 0.291 & \textbf{0.188} \\
        % \hline
        Cotton  & 0.249 & 0.315  & 0.209 & \textbf{0.177} \\
        \hline        
        % Dino Overall & 0.124 & 0.161 & {0.128} & \textbf{0.121} \\
        % \hline
        Dino  & 0.117 & 0.130  & \textbf{0.109} & 0.117 \\
        \hline
        % Sideboard Overall & 0.203 & 0.380 & \textbf{0.191} & {0.208}\\
        % \hline
        Sideboard & 0.223 & 0.336  & \textbf{0.206} & 0.221 \\
        \hline        
        % \textbf{Average Overall} & 0.241 & 0.327 & 0.233 & \textbf{0.209} \\
        % \hline
        \textbf{Average } & 0.234 & 0.307 & \textbf{0.208} & \textbf{0.208}  \\
        \hline
    \end{tabular}
    \caption{RMSE values for 17 views for each scene with the best $\alpha$ value selected over all of the scenes.}
    \vspace{-2mm}
    \label{tab:HonauerTrainingResults}
\end{table}

\setlength{\tabcolsep}{5pt}
\begin{table}[htb]
% \vspace{-2.5mm}
    \centering
    \begin{tabular}{|c | c c c c c|}
        \thickhline
        % \textbf{Method} & Stereo & 7 Views & 12 Views  & 17 Views \\
        % \thickhline
        % $L^2$-$L^2$ & 0.307 & 0.237 & 0.229 & 0.234 \\
        % \hline
        % $L^2$-$L^1$  & 0.530 & 0.342  & 0.293 & 0.307 \\
        % \hline        
        % $L^1$-$L^1$  & 0.397 & 0.221  & 0.209 & \textbf{0.208} \\
        % \hline
        % Welsch-$L^1$ & \textbf{0.249} & \textbf{0.203} & \textbf{0.206} & \textbf{0.208} \\
        % \hline
        \textbf{Method} & Stereo & 5 views & 9 views & 13 views  & 17 views \\
        \thickhline
        $L^2$-$L^2$ & 0.307 & 0.245 & 0.228 & 0.228 & 0.234 \\
        \hline
        $L^2$-$L^1$ & 0.530  & 0.384 & 0.314  & 0.285 & 0.307 \\
        \hline        
        $L^1$-$L^1$ & 0.397  & 0.238 & 0.214  & 0.207 & \textbf{0.208} \\
        \hline
        Welsch-$L^1$ & \textbf{0.249} & \textbf{0.210} & \textbf{0.200} & \textbf{0.203} & \textbf{0.208} \\
        \hline
    \end{tabular}
    \caption{RMSE values averaged over all four scenes for different numbers of views (including the reference view). We choose the $\alpha$ value that gives the best 17 view RMSE value on average.}
    \vspace{-6mm}
    \label{tab:HonauerTrainingViews}
\end{table}

With 17 views, the $L^1$-$L^1$ and the Welsch-$L^1$ algorithms have the same average RMSE (see Table \ref{tab:HonauerTrainingResults}) which is better than the other methods. The key advantage of the Welsch-$L^1$ algorithm in this dataset is that it is more consistent over a greater range of views for a given $\alpha$ value than the other algorithms tested. It has the best RMSE for all numbers of views shown in Table \ref{tab:HonauerTrainingViews}, reaffirming the results we find in our synthetic dataset.

This observed lack of consistency in the $L^1$ results may be caused by the $L^1$ estimator's breakdown point of 50\% \cite{lopuhaa1991breakdown}. With large numbers of views some areas may be occluded in around 50\% of view pairs. Therefore the $L^1$ estimator will be almost at the breakdown point before considering any noise or other non-idealities in these areas. Additionally, with small numbers of views, the areas may be occluded in a greater proportion of view pairs, further contributing to the inconsistency in the $L^1$-$L^1$ results. The Welsch loss function does not have a breakdown point because infinite distances are not possible in a Welsch metric space \cite{Davies2007} and this better limits the influence of these problematic areas on disparity estimation. 
% \vspace{-2mm}

\section{Conclusion}\label{sec:Conclusion}

In this paper we experimentally show that occlusions present a significant problem for multiview disparity estimation algorithms. These occlusions can cause some algorithms, such as those that use an $L^2$ norm for the data-term to perform worse when more views are used. We propose a Welsch-$L^1$ disparity estimation algorithm as a possible solution to this problem. We also define a disciplined warping strategy and method for progressively including views which can be used to avoid removing high frequency content from views. Even in cases with very high numbers of views we show that these approaches produce excellent results both on our synthetic dataset and a 4D Light Field Dataset \cite{Honauer2016} when compared with the $L^2$-$L^2$, $L^2$-$L^1$ and $L^1$-$L^1$ approaches.

\vfill\pagebreak

%\vfill
%\pagebreak

\bibliographystyle{IEEEbib}
\bibliography{strings,refs}

\end{document}